\documentclass{acm_proc_article-sp} 

\usepackage[hyphens]{url}
\usepackage{graphicx}
\usepackage{booktabs}
\usepackage{rotating}

\begin{document}

\title{{GEMMbench}: a framework for reproducible and collaborative benchmarking
of matrix multiplication}
\subtitle{\LARGE \url{https://github.com/dividiti/gemmbench}}

\numberofauthors{1}
\author{
\alignauthor
  Anton Lokhmotov, {\Large \tt dividiti}\\
  \email{\Large \url{anton@dividiti.com}}\\
  \affaddr{ideaSpace West}\\
  \affaddr{3 Charles Babbage Road}\\
  \affaddr{Cambridge, CB3 0GT}\\
  \affaddr{United Kingdom}\\
}

\maketitle

\begin{abstract}

The generic matrix-matrix multiplication (GEMM) is arguably the most popular
computational kernel of the 20th century.
Yet, surprisingly, no common methodology for evaluating GEMM performance
has been established over the many decades of using GEMM for comparing
architectures, compilers and ninja-class programmers.

We introduce GEMMbench, a framework and methodology for evaluating performance
of GEMM implementations.
GEMMbench is implemented on top of Collective Knowledge (CK), a lightweight
framework for reproducible and collaborative R\&D in computer systems.
Using CK allows the R\&D community to crowdsource hand-written and
compiler-generated GEMM implementations and to study their performance across
multiple platforms, data sizes and data types.
Our initial implementation supports hand-written OpenCL kernels operating on
matrices consisting of single- and double-precision floating-point values, and
producing single or multiple output elements per work-item (via thread
coarsening and vectorization).

\end{abstract}


\terms{Software}


\section{Motivation}

The generic matrix-matrix multiplication (GEMM) is given by the equation:
\begin{displaymath} C = \alpha A \times B + \beta C \end{displaymath}
\noindent where $A$, $B$ and $C$ are matrices, and $\alpha$ and $\beta$ are
scalars.

GEMM is arguably the most popular computational kernel of the 20th century.
The apparent simplicity of GEMM has haunted generation after generation of
researchers who have evaluated its performance on generation after generation
of computer systems,\footnote{Conveniently, both for researchers and computer
systems a generation means 3--4 years.} while uncovering layer after layer of
its hidden complexity.
For example, discovering the beneficial effects of cache blocking on GEMM
performance~\cite{Lam:1991} has fuelled research on locality optimizations in
compilers for many years.

Yet, surprisingly, no common methodology for evaluating GEMM performance has
been established over the many decades of using this kernel for comparing
architectures, compilers and ninja-class programmers.
Consequently, the reader of a report presenting GEMM results is often left wondering:

\begin{itemize}
\item Was the kernel specialized, for example, to $C = A \times B$? (In other
words, $\alpha=1$ and $\beta=0$.)
\item Which of the data types were used: single precision (SGEMM), double
precision (DGEMM), complex single precision (CGEMM), or complex double
precision (ZGEMM)?
\item Which data layouts were used: normal (N) or transposed (T)?\footnote{For
matrices stored in row-major order, $C = \alpha A \times B^{T} + \beta C$
typically results in better locality, because $B^{T}$ is read row-wise.} If
transposed, did the execution time include the overhead for transposition?
\item Which data shapes were used: square or rectangular? If rectangular, did
the execution time depend on the ratio between the dimensions?
\item Which data sizes were used: small or large?
\item On a system with caches, did `large' result in cache thrashing; did
`small' result in good locality (no thrashing)?
\item On a heterogeneous system equipped with a discrete accelerator, did the
execution time include the overhead for copying the data to the accelerator and
back, or only the kernel execution time?
\item Did the evaluation include power or energy measurements?
\item If a diesel generator was used to get the system running, how many
megaflops per gallon were they
getting?\footnote{\url{http://www.hpcwire.com/2006/06/30/the_new_limits_on_high_performance_computing-1/}}
\item More seriously, have we achieved significant improvements in energy
efficiency of floating-point operations over the last
decade?\footnote{Manufactured in 2005 on a 130 nm process, ClearSpeed's CSX600
processor provided 25 DGEMM Gflops/s in under 10 Watts according to their
marketing materials.}
\item How much human effort and ingenuity was involved in writing the kernel or
in implementing the compiler that generated the kernel?
\item Can we compare the generators, for example, based on polyhedral
compilation~\cite{Beaugnon:2014} and functional expression
rewriting~\cite{Steuwer:2015} in a fair way (including code quality, code
generation time and robustness)?
\item Can we evaluate the generators against ninja-class
programmers~\cite{Goto:2008} or vendor libraries?
\item Have we used all the tricks up our sleeves to get the fastest GEMM
implementation for our hardware and problem at hand?
\item Can we {\em adapt} our GEMM implementations to work well across a range
of architectures, data types, data sizes, etc.?
\end{itemize}

Given that we are discussing GEMM, a simple kernel intended to
give us insights for solving more complex `real-world' problems, it is
essential to start getting some of the answers right to facilitate our learning
and knowledge sharing.

We introduce GEMMbench, a framework and methodology for evaluating performance
of GEMM implementations.
GEMMbench is implemented on top of Collective Knowledge (CK), a lightweight
framework for reproducible and collaborative R\&D in computer systems.%
\footnote{CK is the last in the lineage of frameworks designed by Grigori
Fursin. See \url{http://cknowledge.org}. Its immediate predecessor, Collective
Mind (CM), is described in the following articles~\cite{CollectiveMind,
CollectiveMind2}.}

Our initial implementation supports hand-written OpenCL kernels operating on
matrices consisting of single- and double-precision floating-point values, and
producing single or multiple output elements per work-item (via thread
coarsening and vectorization).

Over time, we plan to involve the community to add further hand-written and
generated kernels (e.g. from \cite{Beaugnon:2014,Steuwer:2015}), and,
importantly, to collectively study the GEMM performance across multiple
platforms, data sizes and data types.

\section{Implementation}
\label{sec:implementation}

%
The GEMMbench framework reads from a JSON file the metadata describing a
kernel.
The JSON file specifies the data type ({\tt S} or {\tt D}), the layout of the
matrices ({\tt N} or {\tt T}), the thread-coarsening configuration ({\tt di}
for the number of rows and {\tt dj} for the number of columns in a block
computed by a single work-item), and so on.

For example, the SGEMM kernel that assumes that $A$ is non-transposed and $B$
is transposed and outputs a single element per work-item:
\begin{verbatim}
kernel void gemm(
    global float const * restrict A,
    global float const * restrict B,
    global float       * restrict C,
    float alpha, float beta, uint n)
{
    const uint j = get_global_id(0);
    const uint i = get_global_id(1);

    float ABij = 0.0f;
    for (uint k = 0; k < n; k += 1)
    {
        ABij += A[i*n + k] * B[j*n + k];
    }
    C[i*n + j] = alpha * ABij + beta * C[i*n + j];
}
\end{verbatim}
is described by the following metadata:
\begin{verbatim}
{
    "name"   : "SGEMM_NT_1x1",
    "file"   : "SGEMM_NT_1x1.cl",
    "type"   : "S",
    "transA" : "N",
    "transB" : "T",
    "dj"     : 1,
    "di"     : 1
}
\end{verbatim}

See further examples in the {\tt dataset} entries of the GEMMbench repository.%
\footnote{\url{https://github.com/dividiti/gemmbench/tree/master/dataset}}

\section{Installation}
\label{sec:installation}

\subsection{Install Collective Knowledge}

Install Collective Knowledge (CK) {\em e.g.}:\footnote{See
\url{http://github.com/ctuning/ck} for alternative instructions.}
\begin{verbatim}
$ export CK_REPOS=~/CK
$ mkdir $CK_REPOS
$ export CK_ROOT=$CK_REPOS/ck
$ git clone https://github.com/ctuning/ck.git $CK_ROOT
$ export PATH=$CK_ROOT/bin:$PATH
$ ck status
Your version is up-to-date: V1.6.12
\end{verbatim}

For using CK in Python scripts ({\em e.g.}\ with IPython Notebook):
\begin{verbatim}
$ cd $CK_ROOT && sudo python setup.py install
$ python -c "import ck.kernel as ck;
             print ck.version({})['version_str']"
1.6.12
\end{verbatim}
%

\subsection{Install GEMMbench}

Install GEMMbench along with other CK repositories it depends upon: {\tt
ck-autotuning}, {\tt ck-env}, {\em etc.}
\begin{verbatim}
$ ck pull repo:gemmbench \
  --url=https://github.com/dividiti/gemmbench
\end{verbatim}
%

\subsection{Register OpenCL driver}

Register an (already installed) OpenCL driver {\em e.g.}:
\begin{verbatim}
$ ck find soft:lib.opencl*
~/CK/ck-env/soft/lib.opencl.linux
...
$ ck setup soft:lib.opencl.linux
\end{verbatim}

Enter the OpenCL driver version (a string to identify it later). Enter the path
to \verb|libOpenCL.so| without \verb|lib| {\em e.g.} \verb|/usr| for
\verb|libOpenCL.so| in \verb|/usr/lib|.

\subsection{Customize platform scripts}

Take a look at one of the available scripts for disabling frequency and voltage
scaling (DVFS) and fixing the frequencies:
\begin{verbatim}
$ ck find platform.init:*
~/CK/ck-autotuning/platform.init/generic-android
~/CK/ck-autotuning/platform.init/chromebook-ubuntu
~/CK/ck-autotuning/platform.init/generic-odroid
~/CK/ck-autotuning/platform.init/generic-linux
\end{verbatim}

Customize the scripts called from {\tt ck-set-performance} (you may leave them
blank initially) and add them to the system path {\em e.g.}:
\begin{verbatim}
$ cp ~/CK/ck-autotuning/platform.init/generic-linux \
  ~/CK/ck-autotuning/platform.init/my-linux-platform
...
$ export PATH=$PATH:\
  ~/CK/ck-autotuning/platform.init/my-linux-platform
\end{verbatim}
(Ensure the scripts are executable.)

\subsection{Compile GEMMbench}

To compile GEMMbench:
\begin{verbatim}
$ ck compile program:gemmbench-cl-launcher-1.0
\end{verbatim}
(The cJSON and xOpenME libraries should be installed automatically the first
time you compile GEMMbench.)

\subsection{Run GEMMbench}

To run GEMMbench with the default parameters:
\begin{verbatim}
$ ck run program:gemmbench-cl-launcher-1.0
\end{verbatim}

Select the default command (press ``Enter''), one of the four currently
supported ``flavours'' (SGEMM NN, SGEMM NT, DGEMM NN, DGEMM NT), and finally
one of the kernel variants.

To override the default parameters, use {\em e.g.}
\begin{verbatim}
$ ck run program:gemmbench-cl-launcher-1.0 \
  --extra_run_cmd="-p 1 -d 1 -n 512 -lws 4,16"
\end{verbatim}
to run on platform 1, device 1, with the matrix order of 512 and
the local work size of {\tt (4,16)}.

\subsection{Run SGEMM experiments}

\begin{verbatim}
$ ck find script:SGEMM*
~/CK/gemmbench/script/SGEMM_NT
$ cd ~/CK/gemmbench/script/SGEMM_NT
$ ./_clean_program_pipeline.sh
$ ./_setup_program_pipeline.sh
...
Pipeline is ready!
$ ./explore-f-n.sh
$ ./explore-n-lws.sh
\end{verbatim}

\subsection{How to reproduce?}
\label{sec:reproduce}

To replay our experiments, obtain our experimental data for this paper:

\begin{verbatim}
$ ck pull repo:gemmbench-adapt16 \
  --url=https://github.com/dividiti/gemmbench-adapt16
$ ck find experiment:SGEMM_NT*
~/CK/gemmbench-adapt16/experiment/SGEMM_NT-explore-f-n
~/CK/gemmbench-adapt16/experiment/SGEMM_NT-explore-n-lws
\end{verbatim}

Start the CK web server:
\begin{verbatim}
$ ck start web
\end{verbatim}

Open \url{http://localhost:3344} in a web browser.
Select {\tt gemmbench-adapt16} in the ``Repository'' dropdown menu.
Open {\tt SGEMM\_NT-explore-f-n} or {\tt SGEMM\_NT-explore-n-lws}.%
\footnote{You can also go directly to
\url{http://localhost:3344/?wcid=bc0409fb61f0aa82:8bcbe025bd8803c2} or
\url{http://localhost:3344/?wcid=bc0409fb61f0aa82:a697f73cf4392f23}.}
Select {\tt gemmbench-view} in the ``Select experiment view'' menu under a QR
code.

The table shows one experiment (with a number of statistical repetitions) per
row.
Click on a ``Copy to clipboard'' button in the rightmost column to obtain a
command to replay the corresponding experiment.%
\footnote{Section~\ref{sec:evaluation} explains the meaning of most other
columns.}
For example, to replay the intermittently failing experiment mentioned in
Section~\ref{sec:validating}, run:
\begin{verbatim}
$ ck replay \
  experiment:8bcbe025bd8803c2 --point=ca4a5dbe25613c7d
\end{verbatim}
%

\section{Evaluation}
\label{sec:evaluation}

We demonstrate using the GEMMbench framework for evaluating 3 SGEMM NT
OpenCL kernels on a Hardkernel Odroid~XU3 board (Table~\ref{Odroid}).%
\footnote{\url{http://www.hardkernel.com/main/products/prdt_info.php?g_code=G140448267127}}

\subsection{Evaluation platform}

The Odroid XU3 board has 4 integrated power consumption sensors:
\begin{itemize}

  \item for the LPDDR3 RAM;

  \item for the CPU cluster 0 comprised of 4 ARM Cortex-A7 (``LITTLE'') cores;

  \item for the CPU cluster 1 comprised of 4 ARM Cortex-A15 (``big'') cores;

  \item for both the GPU cluster 0 and the GPU cluster 1 comprised respectively
  of 4 and 2 ARM Mali-T628 cores.

\end{itemize}
We reused the {\em pipeline} functionality of the underlying Collective
Knowledge framework to conduct experiments under controlled conditions.
We ran the SGEMM kernels on the GPU cluster 0 (OpenCL device 0); we disabled
dynamic voltage and frequency scaling (DVFS) and set the frequency to the
maximum of 600 MHz.
Likewise, we set the CPU governors to the performance mode and the CPU
frequencies to the maximum.
%

\subsection{OpenCL kernels}

The SGEMM NT kernels are contained in 3 separate files:
\begin{itemize}

  \item \verb|SGEMM_NT_1x1.cl|: a na\"ive version shown in
  Section~\ref{sec:implementation} which computes a single element of matrix $C$ per
  work-item;

  \item \verb|SGEMM_NT_4x1.cl|: a vectorised version which computes a vector of
  four adjacent elements of matrix $C$ per work-item;

  \item \verb|SGEMM_NT_4x1_barrier.cl|: a similarly vectorised version which
  synchronises work-items in a work-group with a barrier to improve cache
  utilisation~\cite{Gronqvist:2014}.%
  \footnote{\url{http://malideveloper.arm.com/downloads/GPU_Pro_5/GronqvistLokhmotov_white_paper.pdf}}

\end{itemize}

\begin{table*}
\centering
\caption{\label{Odroid}Experimental platform: Hardkernel Odroid~XU3 board.}
  \begin{tabular}{ll}
  \toprule
  {\bf Hardware}   &\\
  \midrule
  System-on-chip (SoC)         & Samsung Exynos 5422                     \\
  GPU cluster 0 (``device 0'') & ARM Mali-T628,  4 cores, $\le$ 600 MHz  \\
  GPU cluster 1 (``device 1'') & ARM Mali-T628,  2 cores, $\le$ 600 MHz  \\
  CPU cluster 0 (``LITTLE'')   & ARM Cortex-A7,  4 cores, $\le$ 1400 MHz \\
  CPU cluster 1 (``big'')      & ARM Cortex-A15, 4 cores, $\le$ 2000 MHz \\
  LPDDR3 RAM                   & 2 GiB, $\le$ 14.9 Gbytes/s              \\
  \midrule
  {\bf Software}   &\\
  \midrule
  Board support package (BSP)  & 2015-02-25                    \\
  Ubuntu Linux                 & 14.04.1 (updated to 14.04.3)  \\
  Linux kernel                 & 3.10.69                       \\
  OpenCL version               & 1.1 Full Profile              \\
  OpenCL driver                & 4.0 (BSP default)             \\
  Host compiler                & Clang++ 3.6                   \\
  \bottomrule
  \end{tabular}
\end{table*}

\subsection{Varying the matrix order}
\label{sec:order}

For the first set of experiments (using the {\tt explore-f-n} script), we
varied the matrix order from 64 to 1024, {\em i.e.}\ performed experiments for
the matrix dimensions ranging from $64 \times 64$ to $1024 \times 1024$.
We fixed the OpenCL local work size (work-group size) to $(8,8)$, which,
depending on the register usage, supports up to four concurrently executing
work-groups per Mali-T628 core~\cite{Gronqvist:2014}.

\begin{figure*}
  \includegraphics[width=\textwidth]{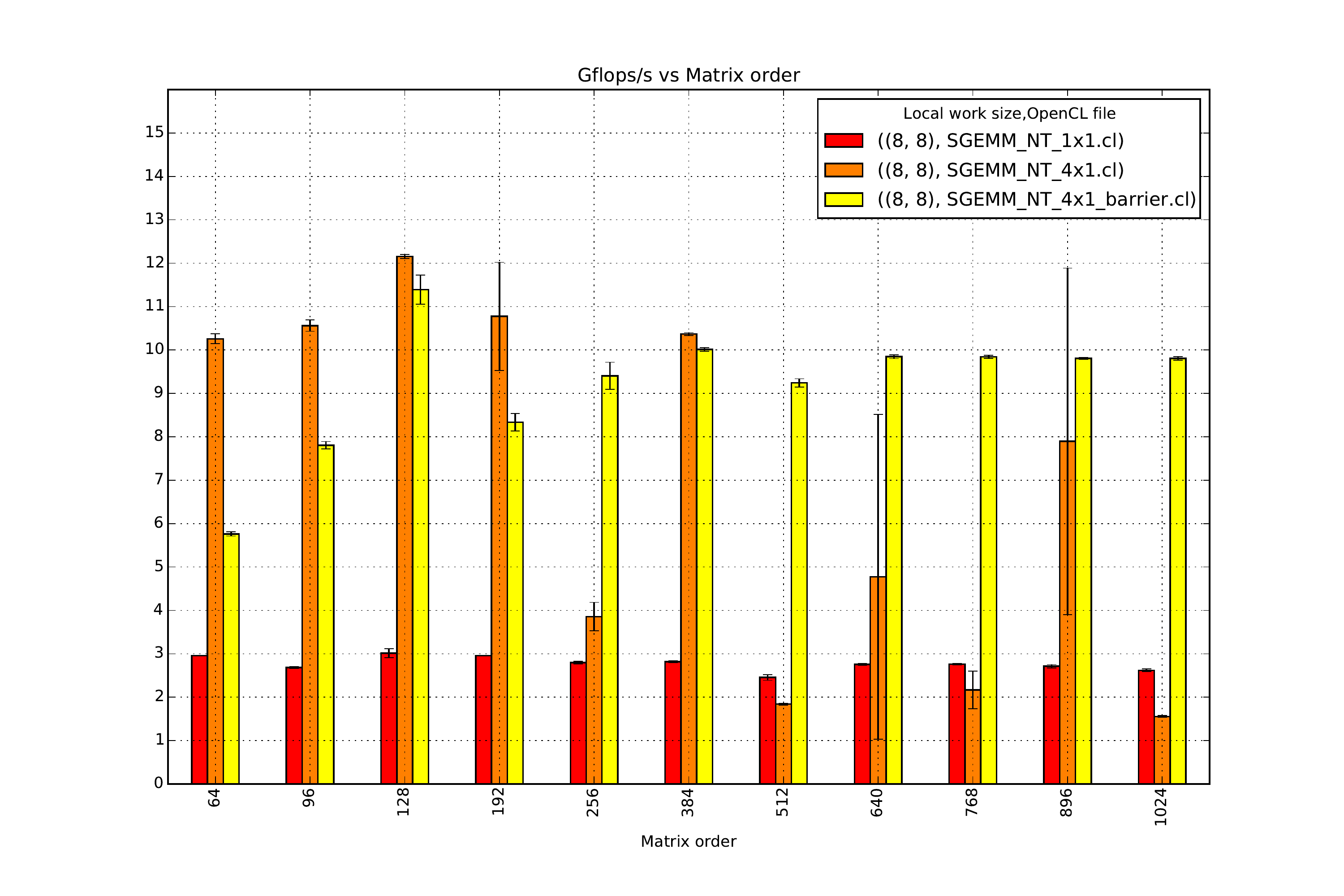}
  \caption{The performance of 3 SGEMM NT kernels.}
  \label{SGEMM_NT:plot}
\end{figure*}

Columns 0--3 of Table~\ref{SGEMM_NT:df} show the raw results in Gflops/s from 4
statistical repetitions (under the same experimental conditions); column 4
shows the mean for each experiment (computed using {\tt pandas.mean()}); column
5 shows the standard deviation (computed using {\tt pandas.std()}).
Figure~\ref{SGEMM_NT:plot} shows the means as a bar plot with the error bars
taken from the standard deviations.

Across the matrix orders, the \verb|SGEMM_NT_1x1.cl| program achieved
low but stable performance up to 3 Gflops/s.
The \verb|SGEMM_NT_4x1.cl| program achieved over 12 Gflops/s but its
performance dropped dramatically for the matrix orders that were a multiple of
256: 256, 512, 768 and 1024.
In addition, it exhibited high performance variation for the matrix orders of
256 and 896.
The \verb|SGEMM_NT_4x1_barrier.cl| program achieved up to 11.5
Gflops/s. Importantly, it maintained high performance of 9--10 Gflops/s
for the matrix orders above 256.

\begin{table*}
  \centering
  \caption{\label{SGEMM_NT:df}The performance of 3 SGEMM NT kernels.}
\begin{tabular}{l|l|l|rrrrrr}
\toprule
{\bf OpenCL program} & {\bf Local work size} & {\bf Matrix order} &       0 &       1 &       2 &       3 &      mean &       std \\
\midrule
SGEMM\_NT\_1x1.cl & (8, 8) & 64   &   2.954 &   2.966 &   2.957 &   2.958 &   2.95875 &  0.005123 \\
                &        & 96   &   2.687 &   2.709 &   2.655 &   2.684 &   2.68375 &  0.022172 \\
                &        & 128  &   3.086 &   2.863 &   3.030 &   3.071 &   3.01250 &  0.102439 \\
                &        & 192  &   2.947 &   2.954 &   2.962 &   2.966 &   2.95725 &  0.008461 \\
                &        & 256  &   2.792 &   2.765 &   2.791 &   2.843 &   2.79775 &  0.032653 \\
                &        & 384  &   2.846 &   2.820 &   2.783 &   2.819 &   2.81700 &  0.025884 \\
                &        & 512  &   2.532 &   2.382 &   2.474 &   2.431 &   2.45475 &  0.063757 \\
                &        & 640  &   2.741 &   2.788 &   2.746 &   2.749 &   2.75600 &  0.021587 \\
                &        & 768  &   2.769 &   2.747 &   2.779 &   2.758 &   2.76325 &  0.013817 \\
                &        & 896  &   2.726 &   2.659 &   2.732 &   2.739 &   2.71400 &  0.037050 \\
                &        & 1024 &   2.663 &   2.591 &   2.614 &   2.606 &   2.61850 &  0.031161 \\
\midrule
SGEMM\_NT\_4x1.cl &        & 64   &  10.393 &  10.207 &  10.130 &  10.299 &  10.25725 &  0.113855 \\
                &        & 96   &  10.726 &  10.596 &  10.429 &  10.494 &  10.56125 &  0.129567 \\
                &        & 128  &  12.115 &  12.139 &  12.141 &  12.225 &  12.15500 &  0.048139 \\
                &        & 192  &   8.952 &  11.737 &  11.070 &  11.340 &  10.77475 &  1.245662 \\
                &        & 256  &   4.038 &   4.002 &   4.026 &   3.363 &   3.85725 &  0.329840 \\
                &        & 384  &  10.412 &  10.350 &  10.373 &  10.339 &  10.36850 &  0.032275 \\
                &        & 512  &   1.863 &   1.821 &   1.824 &   1.858 &   1.84150 &  0.022068 \\
                &        & 640  &   2.932 &  10.388 &   2.886 &   2.886 &   4.77300 &  3.743396 \\
                &        & 768  &   1.982 &   1.916 &   2.818 &   1.960 &   2.16900 &  0.433536 \\
                &        & 896  &  10.337 &  10.336 &   1.990 &   8.916 &   7.89475 &  3.993049 \\
                &        & 1024 &   1.531 &   1.575 &   1.539 &   1.572 &   1.55425 &  0.022500 \\
\midrule
SGEMM\_NT\_4x1\_barrier.cl &        & 64   &   5.742 &   5.746 &   5.833 &   5.728 &   5.76225 &  0.047794 \\
                &        & 96   &   7.680 &   7.840 &   7.867 &   7.835 &   7.80550 &  0.084839 \\
                &        & 128  &  10.969 &  11.587 &  11.722 &  11.280 &  11.38950 &  0.335844 \\
                &        & 192  &   8.580 &   8.180 &   8.413 &   8.165 &   8.33450 &  0.199193 \\
                &        & 256  &   9.053 &   9.263 &   9.772 &   9.530 &   9.40450 &  0.313252 \\
                &        & 384  &  10.075 &   9.997 &   9.991 &   9.989 &  10.01300 &  0.041473 \\
                &        & 512  &   9.327 &   9.263 &   9.272 &   9.109 &   9.24275 &  0.093546 \\
                &        & 640  &   9.895 &   9.875 &   9.822 &   9.798 &   9.84750 &  0.045141 \\
                &        & 768  &   9.817 &   9.869 &   9.882 &   9.809 &   9.84425 &  0.036619 \\
                &        & 896  &   9.840 &   9.781 &   9.806 &   9.800 &   9.80675 &  0.024595 \\
                &        & 1024 &   9.829 &   9.855 &   9.765 &   9.780 &   9.80725 &  0.041955 \\
\bottomrule
\end{tabular}


%
\end{table*}

\subsection{Varying the local work size}
\label{sec:lws}

\begin{sidewaysfigure*}
  \includegraphics[width=\textheight]{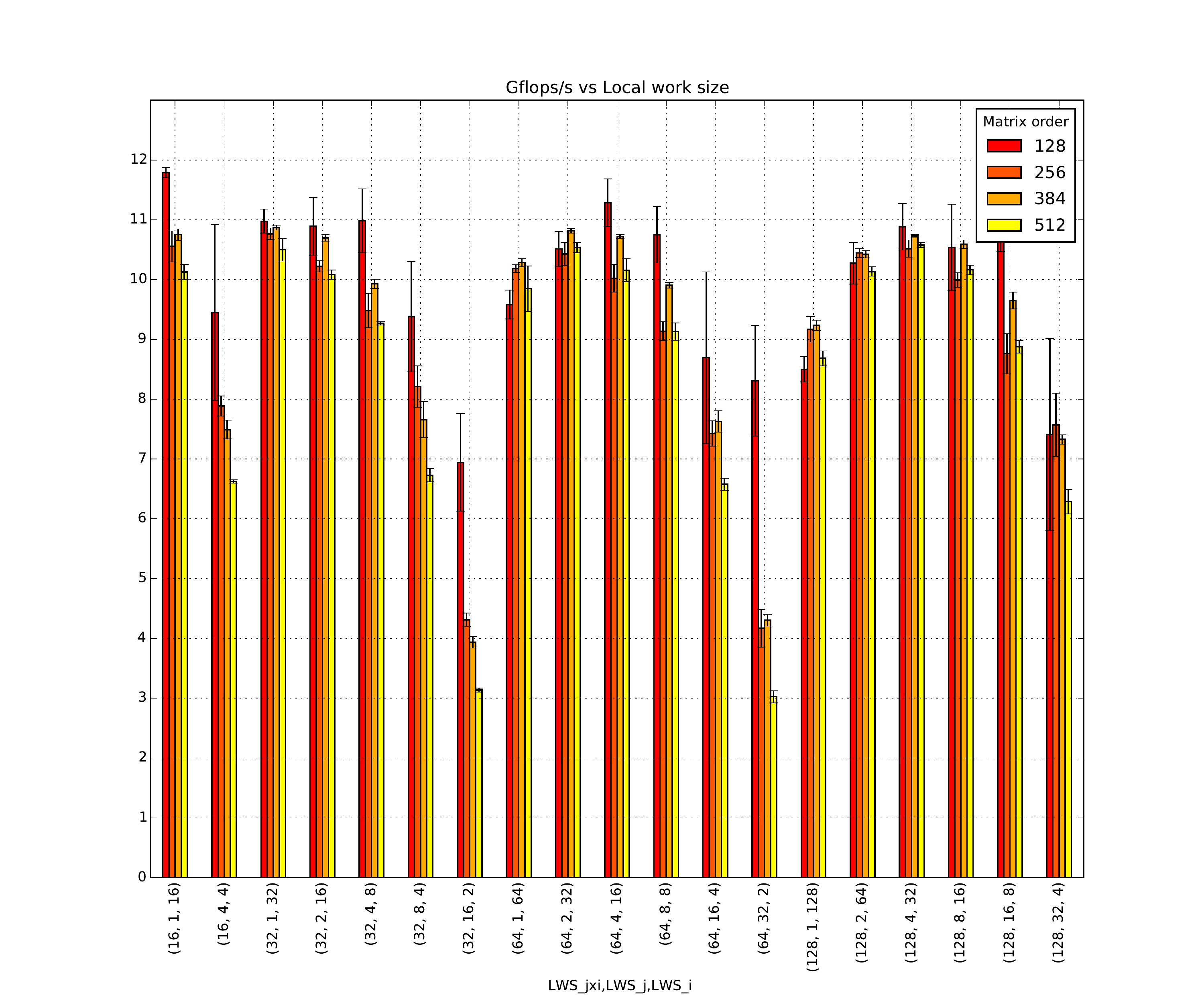}
  \caption{The performance of {\tt SGEMM\_NT\_4x1\_barrier.cl} across work-group sizes with up to 64 work-items.}
  \label{SGEMM_NT:plot:lws}
\end{sidewaysfigure*}

For the second set of experiments (using the {\tt explore-n-lws} script), we
varied the local work size (work-group size) for the
\verb|SGEMM_NT_4x1_barrier.cl| program and 4 values of the matrix order.
Table~\ref{SGEMM_NT:plot:lws} shows a bar plot with the local work size varied
from 16 to 128 work-items per work-group.
For page size limits, Figure~\ref{SGEMM_NT:df:lws} shows the raw data only for
the local work size of up to 64 work-items per work-group.

Overall, by exploring the local work size space we were able to achieve up to
$20\%$ performance improvement over our default of $(8,8)$.
But rather than using exhaustive search we could guide it from a small number
of experiments, as motivated by the following observations.

Initially, we started exploring the local work size space with the first
dimension $j \ge 2$.
We then noticed that using the local work size of $(s_l, s_h)$, where $s_l <
s_h$, was faster than using $(s_h, s_l)$.
For example, using $(2, 16)$ was $1.5--3$ times faster than using $(16, 2)$;
using $(1, 16)$ even resulted in the record $11.9$ Gflops/s for this program.
We run more experiments with $s_l = 1, 2$ and discerned an interesting pattern:
for small local work sizes (16 and 32), we got the best performance with $s_l =
1$; for larger local work sizes (64 and 128), we got the best performance with
$s_l = 4$.
We could use ``predictive analytics'' to discern at least the ``first-order''
effect of the preference for using $(s_l, s_h)$, where $s_l < s_h$, but perhaps
also the ``second-order'' effect.\footnote{This is left as an exercise for the
reader.}

\begin{table*}
  \centering
  \caption{\label{SGEMM_NT:df:lws}The performance of {\tt SGEMM\_NT\_4x1\_barrier.cl} work-group sizes with up to 64 work-items.}
\begin{tabular}{lll|l|rrrrrr}
\toprule
{\bf LWS} $j \times i$ & {\bf LWS} $j$ & {\bf LWS} $i$ & {\bf Order} &       0 &       1 &       2 &       3 &      mean &       std \\
\midrule
16  & 1  & 16  & 128 &  11.833 &  11.689 &  11.880 &  11.754 &  11.78900 &  0.084542 \\
    &    &     & 256 &  10.670 &  10.562 &  10.198 &  10.800 &  10.55750 &  0.258665 \\
    &    &     & 384 &  10.853 &  10.633 &  10.732 &  10.800 &  10.75450 &  0.094940 \\
    &    &     & 512 &  10.035 &  10.103 &  10.311 &  10.067 &  10.12900 &  0.124472 \\
    & 4  & 4   & 128 &   8.486 &  10.376 &  11.008 &   7.944 &   9.45350 &  1.469935 \\
    &    &     & 256 &   7.748 &   7.828 &   8.131 &   7.843 &   7.88750 &  0.167604 \\
    &    &     & 384 &   7.451 &   7.722 &   7.369 &   7.428 &   7.49250 &  0.156849 \\
    &    &     & 512 &   6.648 &   6.653 &   6.597 &   6.603 &   6.62525 &  0.029330 \\
\midrule
32  & 1  & 32  & 128 &  11.225 &  10.983 &  10.969 &  10.737 &  10.97850 &  0.199328 \\
    &    &     & 256 &  10.741 &  10.896 &  10.764 &  10.666 &  10.76675 &  0.095789 \\
    &    &     & 384 &  10.886 &  10.828 &  10.848 &  10.918 &  10.87000 &  0.040033 \\
    &    &     & 512 &  10.604 &  10.511 &  10.657 &  10.233 &  10.50125 &  0.188740 \\
    & 2  & 16  & 128 &  10.183 &  10.987 &  11.149 &  11.251 &  10.89250 &  0.485330 \\
    &    &     & 256 &  10.236 &  10.243 &  10.098 &  10.318 &  10.22375 &  0.091682 \\
    &    &     & 384 &  10.773 &  10.654 &  10.705 &  10.664 &  10.69900 &  0.054043 \\
    &    &     & 512 &  10.111 &   9.992 &  10.170 &  10.063 &  10.08400 &  0.075344 \\
    & 4  & 8   & 128 &  10.424 &  11.690 &  11.045 &  10.785 &  10.98600 &  0.533961 \\
    &    &     & 256 &   9.224 &   9.408 &   9.894 &   9.404 &   9.48250 &  0.287441 \\
    &    &     & 384 &  10.033 &   9.849 &   9.897 &   9.937 &   9.92900 &  0.078111 \\
    &    &     & 512 &   9.290 &   9.284 &   9.234 &   9.268 &   9.26900 &  0.025113 \\
    & 8  & 4   & 128 &   8.766 &  10.441 &   8.472 &   9.847 &   9.38150 &  0.921097 \\
    &    &     & 256 &   8.177 &   7.898 &   8.701 &   8.077 &   8.21325 &  0.345041 \\
    &    &     & 384 &   7.450 &   7.408 &   8.062 &   7.712 &   7.65800 &  0.301051 \\
    &    &     & 512 &   6.638 &   6.886 &   6.680 &   6.714 &   6.72950 &  0.108865 \\
    & 16 & 2   & 128 &   6.986 &   6.005 &   7.991 &   6.799 &   6.94525 &  0.816642 \\
    &    &     & 256 &   4.361 &   4.440 &   4.181 &   4.268 &   4.31250 &  0.112370 \\
    &    &     & 384 &   3.986 &   3.871 &   4.051 &   3.840 &   3.93700 &  0.098593 \\
    &    &     & 512 &   3.145 &   3.181 &   3.101 &   3.118 &   3.13625 &  0.034903 \\
\midrule
64  & 1  & 64  & 128 &   9.883 &   9.635 &   9.516 &   9.302 &   9.58400 &  0.242315 \\
    &    &     & 256 &  10.119 &  10.210 &  10.266 &  10.147 &  10.18550 &  0.065790 \\
    &    &     & 384 &  10.360 &  10.279 &  10.303 &  10.196 &  10.28450 &  0.068081 \\
    &    &     & 512 &  10.047 &   9.286 &   9.982 &  10.081 &   9.84900 &  0.377574 \\
    & 2  & 32  & 128 &  10.674 &  10.077 &  10.607 &  10.692 &  10.51250 &  0.292628 \\
    &    &     & 256 &  10.264 &  10.707 &  10.419 &  10.335 &  10.43125 &  0.194443 \\
    &    &     & 384 &  10.766 &  10.840 &  10.800 &  10.853 &  10.81475 &  0.039559 \\
    &    &     & 512 &  10.426 &  10.548 &  10.636 &  10.535 &  10.53625 &  0.086110 \\
    & 4  & 16  & 128 &  10.968 &  10.927 &  11.710 &  11.538 &  11.28575 &  0.397192 \\
    &    &     & 256 &   9.831 &   9.857 &  10.084 &  10.328 &  10.02500 &  0.231769 \\
    &    &     & 384 &  10.748 &  10.690 &  10.754 &  10.703 &  10.72375 &  0.032004 \\
    &    &     & 512 &  10.057 &  10.341 &   9.944 &  10.293 &  10.15875 &  0.189481 \\
    & 8  & 8   & 128 &  10.113 &  10.789 &  10.852 &  11.248 &  10.75050 &  0.471062 \\
    &    &     & 256 &   9.211 &   9.021 &   8.993 &   9.329 &   9.13850 &  0.159711 \\
    &    &     & 384 &   9.953 &   9.945 &   9.857 &   9.878 &   9.90825 &  0.047940 \\
    &    &     & 512 &   9.253 &   8.939 &   9.097 &   9.237 &   9.13150 &  0.146218 \\
    & 16 & 4   & 128 &  10.681 &   8.777 &   7.442 &   7.878 &   8.69450 &  1.436247 \\
    &    &     & 256 &   7.134 &   7.419 &   7.597 &   7.564 &   7.42850 &  0.211008 \\
    &    &     & 384 &   7.863 &   7.455 &   7.525 &   7.666 &   7.62725 &  0.180004 \\
    &    &     & 512 &   6.591 &   6.447 &   6.691 &   6.584 &   6.57825 &  0.100224 \\
    & 32 & 2   & 128 &   8.237 &   9.574 &   7.350 &   8.075 &   8.30900 &  0.927334 \\
    &    &     & 256 &   4.351 &   3.869 &   4.520 &   3.943 &   4.17075 &  0.314849 \\
    &    &     & 384 &   4.163 &   4.365 &   4.321 &   4.375 &   4.30600 &  0.098177 \\
    &    &     & 512 &   3.075 &   2.934 &   2.944 &   3.145 &   3.02450 &  0.102861 \\
\bottomrule
\end{tabular}


%
\end{table*}

\subsection{Comparing energy consumption}

Using the integrated sensors on the Odroid XU3 board, we estimated energy
consumption for the program region that launches a kernel and waits for its
completion.\footnote{We used a rather crude method of averaging the power
consumption measurements at the start and end of the region and multiplying the
average by the execution time.}

\begin{figure*}
  \includegraphics[width=\textwidth]{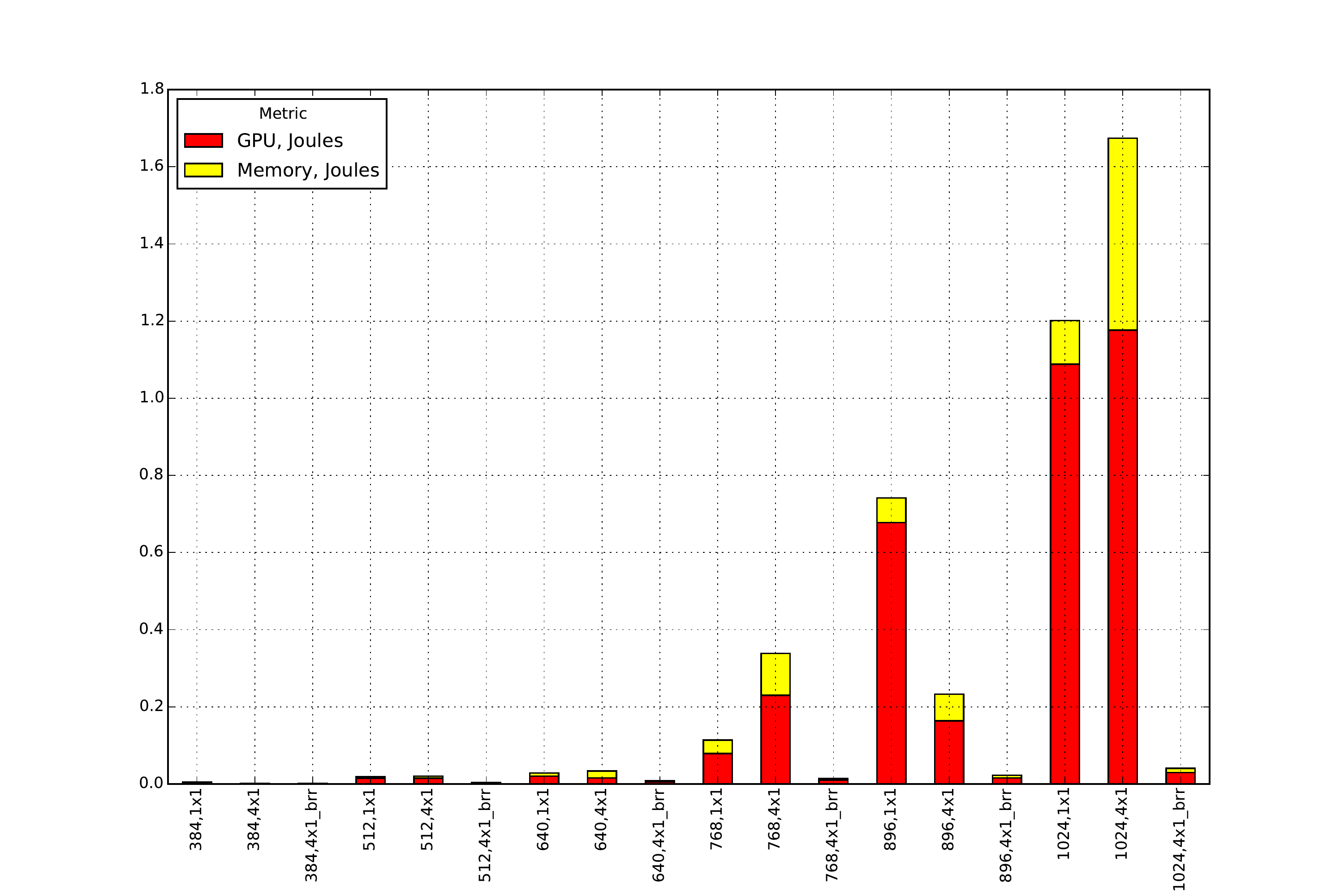}
  \caption{The GPU \& memory energy consumption of 3 SGEMM NT kernels.}
  \label{SGEMM_NT:plot:energy}
\end{figure*}

Table~\ref{SGEMM_NT:df:energy} shows the estimated GPU and memory energy
consumption in Joules across the 3 kernels and 11 matrix orders in our first
set of experiments (Section~\ref{sec:order}).
Figure~\ref{SGEMM_NT:plot:energy} focusses on the energy consumption for the
orders from 384 to 1024.

For the orders of 128 and 384, when the vectorised kernels match in
performance, they also match in energy consumption.
For the order of 1024, however, the non-cache optimised kernel is a disaster:
the cache-optimised vectorised kernel is $6$ times faster and $40$ times more
energy efficient both for the GPU and the RAM;
even the non-vectorised kernel is $75\%$ faster, $10\%$ more energy efficient
for the GPU and $5$ times for the RAM.

\begin{table*}
  \centering
  \caption{\label{SGEMM_NT:df:energy}The GPU \& memory energy consumption of 3 SGEMM NT kernels.}
\begin{tabular}{l|rr|rr|rr}
\toprule
{\bf OpenCL program} & \multicolumn{2}{c}{SGEMM\_NT\_1x1.cl} & \multicolumn{2}{c}{SGEMM\_NT\_4x1.cl} & \multicolumn{2}{c}{SGEMM\_NT\_4x1\_barrier.cl} \\
\midrule
{\bf Metric} &     GPU, Joules & Memory, Joules &     GPU, Joules & Memory, Joules &             GPU, Joules & Memory, Joules \\
\midrule
{\bf Matrix order} &                 &                &                 &                &                         &                \\
64           &        0.000171 &       0.000051 &        0.000151 &       0.000044 &                0.000155 &       0.000046 \\
96           &        0.000225 &       0.000066 &        0.000162 &       0.000048 &                0.000169 &       0.000050 \\
128          &        0.000299 &       0.000086 &        0.000178 &       0.000054 &                0.000183 &       0.000055 \\
192          &        0.000736 &       0.000216 &        0.000287 &       0.000107 &                0.000346 &       0.000122 \\
256          &        0.001643 &       0.000419 &        0.001052 &       0.000403 &                0.000520 &       0.000172 \\
384          &        0.004275 &       0.001199 &        0.001281 &       0.000360 &                0.001325 &       0.000372 \\
512          &        0.014602 &       0.003964 &        0.015229 &       0.005248 &                0.003184 &       0.001125 \\
640          &        0.020130 &       0.008426 &        0.015811 &       0.017664 &                0.005807 &       0.002202 \\
768          &        0.078699 &       0.034993 &        0.230128 &       0.108379 &                0.010011 &       0.004053 \\
896          &        0.677710 &       0.063592 &        0.163536 &       0.069416 &                0.015879 &       0.006794 \\
1024         &        1.088290 &       0.113272 &        1.176622 &       0.497578 &                0.029643 &       0.011166 \\
\bottomrule
\end{tabular}


\end{table*}

\subsection{Validating the results}
\label{sec:validating}

No benchmark should be complete without checking the results for correctness.
GEMMbench includes a reference CPU implementation that an OpenCL
implementation's results are compared against.
Since GEMM operates on floating-point data, the results cannot be expected to be
bit-exact.
Instead, the results are compared element-wise using a small {\em epsilon}
value $\epsilon$.

Initially, we chose $\epsilon = 10^{-5}$.
Early in the development, we accidentally used the integer \verb|abs()|
function instead of the floating-point \verb|fabs()| function.
As a result, element-wise discrepancies not exceeding $1.0$ would frequently go
unnoticed.
Once that issue was fixed, we realised that $\epsilon = 10^{-5}$ was too small
a value when operating on single-precision floating-point data.
Empirically, we found that $\epsilon = 0.1$ worked well in practice when the
elements of the input matrices were drawn from the uniform distribution over
the range $(-0.5, +0.5)$.

Table~\ref{SGEMM_NT:df:match} shows the maximum absolute difference found via
the element-wise comparison and whether the results match under the chosen
$\epsilon = 0.1$ for our first set of experiments (Section~\ref{sec:order}).

\begin{sidewaystable*}
  \centering
  \caption{\label{SGEMM_NT:df:match}The validation of 3 SGEMM NT kernels: {\tt pandas} DataFrame with raw results.}
\begin{tabular}{l|l|rr|rr|rr|rr}
\toprule
                &      &  Max abs diff &  Match? &  Max abs diff &  Match? &  Max abs diff &  Match? &  Max abs diff &  Match? \\
\midrule
SGEMM\_NT\_1x1.cl & 64   &  4.694088e+09 &       0 &  4.694088e+09 &       0 &  2.079083e+05 &       0 &  8.029115e+31 &       0 \\
                & 96   &  5.013380e-02 &       1 &  5.013380e-02 &       1 &  5.013380e-02 &       1 &  5.013380e-02 &       1 \\
                & 128  &  5.014443e-02 &       1 &  5.014443e-02 &       1 &  5.014443e-02 &       1 &  5.014443e-02 &       1 \\
                & 192  &  5.016088e-02 &       1 &  5.016088e-02 &       1 &  5.016088e-02 &       1 &  5.016088e-02 &       1 \\
                & 256  &  5.016087e-02 &       1 &  5.016087e-02 &       1 &  5.016087e-02 &       1 &  5.016087e-02 &       1 \\
                & 384  &  5.016086e-02 &       1 &  5.016086e-02 &       1 &  5.016086e-02 &       1 &  5.016086e-02 &       1 \\
                & 512  &  5.016118e-02 &       1 &  5.016118e-02 &       1 &  5.016118e-02 &       1 &  5.016118e-02 &       1 \\
                & 640  &  5.016124e-02 &       1 &  5.016124e-02 &       1 &  5.016124e-02 &       1 &  5.016124e-02 &       1 \\
                & 768  &  5.016124e-02 &       1 &  5.016124e-02 &       1 &  5.016124e-02 &       1 &  5.016124e-02 &       1 \\
                & 896  &  5.016124e-02 &       1 &  5.016124e-02 &       1 &  5.016124e-02 &       1 &  5.016124e-02 &       1 \\
                & 1024 &  5.016124e-02 &       1 &  5.016124e-02 &       1 &  5.016124e-02 &       1 &  5.016124e-02 &       1 \\
\midrule
SGEMM\_NT\_4x1.cl & 64   &  1.780926e+27 &       0 &  5.013123e-02 &       1 &  8.069804e+19 &       0 &  5.013123e-02 &       1 \\
                & 96   &  8.397023e-02 &       1 &  7.735449e-02 &       1 &  9.320402e-02 &       1 &  7.827297e-02 &       1 \\
                & 128  &  5.014434e-02 &       1 &  5.014434e-02 &       1 &  5.872607e-02 &       1 &  7.191819e-02 &       1 \\
                & 192  &  5.016094e-02 &       1 &  5.016094e-02 &       1 &  5.016094e-02 &       1 &  5.016094e-02 &       1 \\
                & 256  &  5.016080e-02 &       1 &  5.016080e-02 &       1 &  5.016080e-02 &       1 &  5.016080e-02 &       1 \\
                & 384  &  5.016074e-02 &       1 &  5.016074e-02 &       1 &  5.016074e-02 &       1 &  5.016074e-02 &       1 \\
                & 512  &  5.016108e-02 &       1 &  5.016108e-02 &       1 &  5.016108e-02 &       1 &  5.016108e-02 &       1 \\
                & 640  &  5.016237e-02 &       1 &  5.016237e-02 &       1 &  5.016237e-02 &       1 &  5.016237e-02 &       1 \\
                & 768  &  5.016196e-02 &       1 &  5.016196e-02 &       1 &  5.016196e-02 &       1 &  5.016196e-02 &       1 \\
                & 896  &  5.016166e-02 &       1 &  5.016166e-02 &       1 &  5.016166e-02 &       1 &  5.016166e-02 &       1 \\
                & 1024 &  5.016184e-02 &       1 &  5.016184e-02 &       1 &  5.016184e-02 &       1 &  5.016184e-02 &       1 \\
\midrule
SGEMM\_NT\_4x1\_barrier.cl & 64   &  1.390356e-01 &       0 &  1.074878e-01 &       0 &  1.390356e-01 &       0 &  1.390356e-01 &       0 \\
                & 96   &  1.927244e-01 &       0 &  1.927244e-01 &       0 &  1.927244e-01 &       0 &  1.927244e-01 &       0 \\
                & 128  &  9.780534e-02 &       1 &  7.548014e-02 &       1 &  7.864504e-02 &       1 &  8.155191e-02 &       1 \\
                & 192  &  5.630336e-01 &       0 &  5.630336e-01 &       0 &  5.630336e-01 &       0 &  5.630336e-01 &       0 \\
                & 256  &  5.016080e-02 &       1 &  5.016080e-02 &       1 &  5.016080e-02 &       1 &  5.016080e-02 &       1 \\
                & 384  &  5.016074e-02 &       1 &  5.016074e-02 &       1 &  5.016074e-02 &       1 &  5.016074e-02 &       1 \\
                & 512  &  5.016108e-02 &       1 &  5.016108e-02 &       1 &  5.016108e-02 &       1 &  5.016108e-02 &       1 \\
                & 640  &  5.016237e-02 &       1 &  5.016237e-02 &       1 &  5.016237e-02 &       1 &  5.016237e-02 &       1 \\
                & 768  &  5.016196e-02 &       1 &  5.016196e-02 &       1 &  5.016196e-02 &       1 &  5.016196e-02 &       1 \\
                & 896  &  5.016166e-02 &       1 &  5.016166e-02 &       1 &  5.016166e-02 &       1 &  5.016166e-02 &       1 \\
                & 1024 &  5.016184e-02 &       1 &  5.016184e-02 &       1 &  5.016184e-02 &       1 &  5.016184e-02 &       1 \\
\bottomrule
\end{tabular}


%
\end{sidewaystable*}

For the \verb|SGEMM_NT_1x1.cl| program and the matrix order of 64, no results
matched.
In fact, the maximum absolute differences suggest a possible bug in either the
OpenCL or the reference implementation. (Neither has been around for long or
code-reviewed.)

For the \verb|SGEMM_NT_4x1.cl| program and the matrix order of 64, the results
are mixed: they twice matched and twice did not.
The failures may be difficult to debug, since they are intermittent.
Luckily, replaying an experimental point under the Collective Knowledge
framework is a matter of running a single command\footnote{See the end of
Section~\ref{sec:reproduce} for the command to replay this very experiment.}
which would help investigate the failures and quickly test potential solutions.

For the \verb|SGEMM_NT_4x1_barrier.cl| program, the results show repeatable
failures for the orders of 64, 96, 192.
This may give us a clue to what goes wrong here.

Returning to choosing the value of $\epsilon$, we note that these results would pass
under $\epsilon = 0.2$.
It is likely, however, that even this $\epsilon$ would need to be changed had
the input values been drawn from a different distribution {\em e.g.}\ the
uniform distribution over the range $(-5.0, +5.0)$.

Rather than making an arbitrary choice of $\epsilon$ for an arbitrary choice of
the random distribution, we could look into defining representative datasets.
Ideally, they would come from real-world problems along with precision
requirements.
%

\section{Conclusion and outlook}

We have presented GEMMbench, a framework and methodology for systematically
evaluating performance of matrix multiplication implementations.
Our initial implementation supports hand-written OpenCL kernels, producing
single or multiple output elements per work-item (via thread coarsening and
vectorization).

Our goal is to involve the community to extend GEMMbench to evaluate
performance of compiler-generated OpenCL kernels, non-OpenCL implementations,
library implementations and so on, across many target platforms.
To this end, the underlying Collective Knowledge framework provides unique
opportunities for the community to gradually gather and share valuable
knowledge for optimizing performance of matrix multiplication and other
programs, as well as of compilers and processors.
We will build upon other strengths of the Collective Knowledge framework including
support for multiple operating systems (Windows, Linux, Android and MacOS),
compilers (LLVM, GCC, ICC, MSVC, {\em etc.}) and interfaces to packages for
data mining and predictive analytics.

Where do we start?
First, we encourage the interested reader help us investigate the failures
reported in Section~\ref{sec:validating}.
Eric S.\ Raymond's proposition that ``given enough eyeballs, all bugs are
shallow''\footnote{\url{https://en.wikipedia.org/wiki/The_Cathedral_and_the_Bazaar}}
should apply well in this case.
Indeed, some of the failures may have nothing to do with numerical
(in)stability.
Programming errors (such as tacit assumptions) may be detected with static and
dynamic analysis tools such as GPUVerify~\cite{GPUVerify} and
Oclgrind~\cite{Oclgrind}.
In the final version of this article, we will acknowledge those who help us
explain the existing failures and perhaps find new ones, and of course fix them.

Second, we welcome contributions in the form of experimental data in the
Collective Knowledge format.
The contributors should acknowledge their compliance with the terms of use of
their systems, specifically that they do not breach confidentiality.%
\footnote{Note that such terms are easy to overlook. For example, from a
vendor's click-through end-user license agreement: ``BENCHMARKING: This Licence
does not prevent you from using the Software for internal benchmarking
purposes. However, you shall treat any and all benchmarking data relating to
the Software, and any other results of your use or testing of the Software
which are indicative of its performance, efficacy, reliability or quality, as
confidential information and you shall not disclose such information to any
third party without the express written permission of VENDOR.'' Similar terms
have prevented us from sharing results from other platforms in
Section~\ref{sec:evaluation}. As far as we are aware, the standard Hardkernel
BSP is distributed without such limitations.}

Third, we welcome contributions in the form of improvements for the core
GEMMbench code {\em e.g.}\ support for rectangular matrices and complex
floating-point numbers.

Fourth, we welcome contributions in the form of OpenCL kernels or other
implementations.
The initial set of kernels optimised for the ARM Mali-T600 architecture is
intentionally small.
We would like to see contributed kernels optimised for different architectures.
The contributors should acknowledge their copyright in the source
code\footnote{No direct copying from vendors' SDKs or programming guides
please.} and specify the licensing terms.\footnote{The standard Collective
Knowledge license (3-clause BSD) is preferred.}

Fifth, we welcome contributions in the form of representative datasets or
their descriptions.

We envision GEMMbench to inspire community-driven development of other
representative workloads for use in performance evaluation and optimisation of
computer systems.

\section{Acknowledgments and more}

We thank Grigori Fursin, CTO of {\tt dividiti} and Chief Scientist of the
cTuning foundation, for designing and implementing the Collective
Knowledge framework, on top of which we implemented GEMMbench.

We look forward to the public discussion of this draft and hope to get useful
feedback and contributions from the community, which will be acknowledged in
the final version of this article.




\appendix

\section{Sharing experimental results}

Sharing GEMMbench experimental results is easy!

First, run a GEMMbench script {\em e.g.}\ {\tt explore-f-n}, creating locally
an experiment entry in \verb|~/CK/local/experiment| {\em e.g.} {\tt
SGEMM\_NT\_explore-f-n}.

Second, create an empty Git repository on GitHub or elsewhere, and pull from
its URL {\em e.g.}\
\begin{verbatim}
$ ck pull repo:gemmbench-new \
  --url=<new repository's URL>
\end{verbatim}

Third, move or copy your experiments from the local repository to the new repository {\em e.g.}\
\begin{verbatim}
$ ck mv experiment:SGEMM_NT_explore-f-n \
  gemmbench-new:experiment:SGEMM_NT_explore-f-n
\end{verbatim}

Optionally, provide a license file and update further details in {\tt
.cm/info.json} and {\tt .cm/meta.json} in your experimental entries.
(If not provided, the standard CK license and copyright details will apply.)

Finally, commit all the files:
\begin{verbatim}
$ cd `ck find repo:gemmbench-new`
$ git add -A
$ git commit -m "New GEMMbench experiments."
\end{verbatim}

Please do not forget to send us a link to your new repository! We will maintain a list
of such repositories for the community to access, similarly to this paper's
repository (Section~\ref{sec:reproduce}).

\balancecolumns

\end{document}